\documentclass[sigconf, nonacm]{acmart}

\usepackage{multicol}
\usepackage{subfig}
\usepackage{pslatex}

\newcommand\systemname{Micr'Olonys}
\renewcommand{\baselinestretch}{0.98}

\begin{document}
\title{Universal Layout Emulation for Long-Term Database Archival}

\author{Raja Appuswamy}
\affiliation{%
  \institution{EURECOM}
  \city{Biot}
  \state{France}
}
\email{raja.appuswamy@eurecom.fr}

\author{Vincent Joguin}
\affiliation{%
  \institution{EUPALIA}
  \city{Hy\`{e}res}
  \state{France}
}
\email{vincent.joguin@eupalia.com}

\begin{abstract}
Research on alternate media technologies, like film, synthetic DNA, and glass, for long-term data archival  has received a lot of attention recently due to the media obsolescence issues faced by contemporary storage media like tape, Hard Disk Drives (HDD), and Solid State Disks (SSD). While researchers have developed novel layout and encoding techniques for archiving databases on these new media types, one key question remains unaddressed: How do we ensure that the decoders developed today will be available and executable by a user who is restoring an archived database several decades later in the future, on a computing platform that potentially does not even exist today? 

In this paper, we make the case for Universal Layout Emulation (ULE), a new approach for future-proof, long-term database archival that advocates archiving decoders together with the data to ensure successful recovery. In order to do so, ULE brings together concepts from Data Management and Digital Preservation communities by using emulation for archiving decoders. In order to show that ULE can be implemented in practice, we present the design and evaluation of \systemname{}, an end-to-end long-term database archival system that can be used to archive databases using visual analog media like film, microform, and archival paper. 

\end{abstract}

\maketitle

\section{Introduction}

Driven by the promise of machine learning and data analytics, enterprises routinely gather vast amounts of data from diverse data sources. Analysts have reported that enterprise data stored in databases, data warehouses, and data lakes, is growing 40\% annually and will account for 60\% of the 160 Zettabyte Global Datasphere by 2025~\cite{seagate-report}. However, not all data is accessed uniformly. Studies have reported that only 20\% of data stored is performance-critical and accessed frequently. The remaining 80\% is \emph{cold} and accessed infrequently~\cite{IntelWhitePaper}. Historic data used for trend forecasting, archival data stored for meeting legal and regulatory audits, and backup data accessed during failures, are examples of such cold data. Cold data has been identified as the fastest growing data segment with a 60\% annual growth rate, and also as the segment with the longest lifetime (window between creation and deletion date) with retention periods lasting 50--60 years~\cite{100year-archive-2017}. Thus, enterprises are in desperate need of cost-effective options for long-term storage of cold data. 


Traditionally, DBMS have used a tiered storage hierarchy composed of a DRAM or SSD-based performance tier, a HDD-based capacity tier, and a tape-based archival tier. Thus, infrequently-accessed archival data was stored using tape, as it has the lowest cost/GB among all commercially available storage technologies. Unfortunately, tape suffers from two fundamental limitations that complicate long-term database archival. First, tape has a limited lifetime of 10 to 20 years. 
In contrast, enterprises routinely archive data for over 50 years in order to meet legal and regulatory compliance requirements~\cite{100year-archive-2017}. Second, tape density improves at an annual rate of 30\%~\cite{fontana-report,cacm-five-minute-rule}, and tape vendors retain backwards compatibility only up to two generations. As a result of these two limitations, using tape for long-term storage mandates periodic, expensive data migration to deal with device failures and technology upgrades. 
These limitations make tape a less-than-ideal medium for long-term archival of cold data in enterprise databases. 

Recently, several new initiatives have emerged from both industry and academia in an effort to develop new long-term storage technologies that can overcome the limitations of tape. Libraries and archives have long used analog media, like microfilm and archival paper, for protecting journals and magazines across several decades~\cite{promicroform}. For instance, LE-500 rated microfilms and ISO 9706 rated archival paper are designed to last 500 years or more when stored in proper conditions. Recently, photographic media (film) has been used for the preservation of the Declaration of Children's Rights document in collaboration with the UN in the Arctic World Archive~\cite{unicef-archive}. Researchers have also demonstrated the use of synthetic DNA~\cite{dna-archive,goldman-dna-storage,oligoarchive} and glass~\cite{silica-hotstrage18} as promising storage media with orders of magnitude higher density, and thousands of years of longevity. While these initiatives address the storage lifetime challenges associated with long-term archival, enterprise DBMS also face another challenge in archiving data--one related to format obsolescence.

All modern DBMS use proprietary layout formats for storing data. These layouts employ sophisticated compression, partitioning, deduplication, and data organization techniques to improve performance and reduce space utilization. As layout formats evolve to provide new functionality, most commercial DBMS maintain backwards compatibility to ensure that an upgrade to a newer DBMS version does not render a database stored in an older format unusable. However, in the context of long-term preservation, data must be stored in a layout format that is forwards compatible with all future versions of DBMS software. 
Due to this reason, the state-of-the-art approach for long-term database archival is to convert data from a machine-readable, high-performance binary layout to a human-readable, textual representation that uses well-established, publicly-available standards like CSV and XML~\cite{chronos,siard}. The typical approach is to use external tools that communicate with the DBMS using well-established interfaces, and ``dump'' a database into a generic text file that is then archived using a long-term storage medium. 

Unfortunately, this approach suffers from two major drawbacks. First, the switch from binary to textual layout leads to severe data bloat as it strips away the benefit of techniques like compression and deduplication that the database can apply using its knowledge of the schema. As databases continue to be extended with support for more complex data types and more data models, this approach also requires continuously refining the definition of a ``standard'' archival layout to accommodate new data types. 
This is less than ideal especially when taken in the context of long-lasting storage media, as data needs to be migrated from one standard to the next periodically,  and a suite of archival tools that provides compatibility across generations of standards needs to be maintained across decades. Second, while the switch to DBMS-agnostic layout solves the format problem at the database level, it does not solve the problem at the media level; the text file generated from a database dump still has to be converted into a ``physical'' layout format suitable for the long-term storage media. For instance, storing a database on film requires encoding it from its digital form, which is a sequence of bits, into an analog form, which is typically a sequence of barcodes. Similarly, storing a database on DNA requires encoding bits into a sequence of DNA strands. While researchers have developed novel media layout techniques for storing data on these new media types, little attention has been paid to the fact that in order to recover data back successfully, the layout decoders and their parameters should also be archived together with the data. 


In this work, we present \systemname{}--an end-to-end, long-term database archival solution that solves all the aforementioned problems by design. At the core of \systemname{} is a new approach to archival we refer to as \emph{Universal Layout Emulation (ULE)}. The central idea behind ULE is to archive data together with the layout decoders necessary for retrieving the data to ensure that data stored in a custom, binary, compressed layout format can be retrieved using any computing environment in the future. To do so, ULE uses emulation to create a software processor with a custom Instruction Set Architecture, and archives layout algorithms by porting them to this ISA. \systemname{} implements ULE by porting both database and media layout decoding algorithms to DynaRisc, a custom 23-ISA software processor. The instruction stream to be emulated, and the DynaRisc emulator itself are stored together with the data. Using a novel nested emulator design, \systemname{} makes it possible for any user in the future to bootstrap the DynaRisc emulator by writing less than 300 lines of code in any programming language, runtime, or operating system that might not even exist today. We show that empowered by ULE, \systemname{} can perform long-term archival of databases across several types of visual analog media including film, microform, and archival paper.




\section{ULE Motivation}


The problem of preserving software in such a way that it can be executed several decades later on an unknown computing platform is not unique to database archival. Libraries and museums have long faced this problem as they need to permanently preserve an increasingly larger collection of born-digital software artefacts and documents that have historic or cultural significance~\cite{sota18-dp}. One state-of-the-art approach used by researchers in Digital Preservation for preserving software is to use emulation~\cite{rothenberg_1998,emulation-nlpaper}. Emulation refers to the technique that enables a host system to run software or use peripheral devices designed for a different guest system. Emulators simulate the processor and associated guest hardware entirely in software by interpreting the instructions of the guest processor. Thus, emulators can run unmodified software compiled for a guest processor architecture on a host processor with a different architecture. Emulation differs from virtualization whose goal is to provide mediated shared access to underlying hardware. Hypervisors implement virtualization and provide virtual machines that can host unmodified guest OS and applications with minimal overhead by directly executing the guest instructions on the host processor, and by exploiting processor specific virtualization extensions. Thus, virtualization is fundamentally tied to the underlying host hardware. In contrast to virtualization which extends the host processing environment to the guest, emulation tries to faithfully reproduce a guest processing environment on a different host. Thus, emulators prioritize portability and compatibility over performance. Due to these reasons, emulation is being actively used today for preserving historic software designed for old, obsolete computing environments by emulating them on modern processing environments. 


One approach of using emulation for database archival is to archive the DBMS software stack and store it together with the data. At restoration time, an emulator can be used to create the right hardware and software environment for emulating the archived DBMS version in order to access the data. Unfortunately, this approach is not feasible in practice as it suffers from several drawbacks. First, this approach requires the entire DBMS software stack, including supporting libraries, runtimes and OS, to be meticulously archived each time data is archived. This is no simple task given that modern DBMS engines are complex pieces of software with many dependencies. Second, archiving DBMS software also implies that each restoration will emulate one specific older version of the DBMS. Thus, the user now has to perform manual synchronization between the archived versions and the latest version. Third, this approach complicates licensing as emulated older versions have to be potentially licensed differently from non-emulated current versions. Fourth, and perhaps most importantly, the use of emulation for archiving DBMS software simply shifts the problem, as it implicitly assumes the existence of an emulator in the future that can faithfully reproduce the computing environment for the DBMS. As modern DBMS engines are advanced pieces of software that often use processor specific extensions for accelerating performance, they would require emulators to continuously keep pace with advances in instruction set extensions like vector extensions for SIMD, transactional extensions for Hardware Transactional Memory, etcetera, for every architecture supported by the DBMS. This is clearly a non-trivial, non-scalable endeavor.

The ULE approach avoids these problems by not emulating the DBMS software and instead, only emulating the decoders necessary for retrieving data. Such an emulator does not need to emulate a complex architecture like x86 as the goal of the emulator is not to faithfully reproduce unmodified, existing x86 software. Rather, the goal is to be able to simply archive the decoding logic for later execution. Thus, rather than designing an emulator for a given architecture determined by software, we can write the decoding software to target a pre-designed emulator. Such an emulator can simulate a much simpler RISC processor with a limited instruction set that is sufficient for implementing decoders. Note that the processor emulated no longer needs to correspond to any real processor. Thus, the emulator here functions in principle like an interpreter that reads instructions corresponding to the program and interprets them. We refer to such a strategy as \emph{Universal Emulation} to highlight the hardware and architecture-agnostic nature of this approach, and to distinguish it from traditional hardware emulation. Universal Emulation was originally proposed as an approach for long-term software preservation in digital libraries~\cite{uvc} and has not been used for database archival.

Universal Emulation has several advantages for database archival. First, the emulator itself is dramatically simple compared to a traditional hardware emulator due to the limited ISA of the virtual processor. Further, the ISA is a fixed interface that will never be extended unlike current processors that expose new functionality via ISA extensions. On those grounds, there is no reason to continuously maintain and port the emulator across years. In fact, decoders can be archived by simply archiving their instruction stream together with a description of the fixed ISA they are programmed against. During restoration, the ISA description can be used to implement the Universal Emulator using any programming language, OS, or runtime, and the decoder can be executed using the simulated virtual processor. 
The second advantage of ULE is that it seamlessly extends the current archival infrastructure. During archival, layout encoders can compress the textual database archive using database-specific binary layouts and transform them for storage using media-specific layouts. During restoration, the decoders are executed by the emulator to convert the data back into a software-independent format. Thus, ULE uses the same interfaces as traditional archiving for getting data into and out of a DBMS. But, by using universal emulation of decoders, ULE enables the use of structure-aware, media-specific layouts for archiving databases efficiently using long-term storage media. Third, ULE obviates the need for emulating a full DBMS. Thus, queries can be executed at bare-metal performance without any overhead. There is also no need to synchronize data across multiple versions. 
\section{\systemname{} Design}

\systemname{} is a ULE-based archival system we have developed for archiving databases using analog media, like film, microform, and archival paper. In this section, we will explain the design of the three fundamental building blocks of \systemname{}, namely, \emph{DBCoder}, the database layout encoder/decoder, \emph{MOCoder}, the media layout encoder/decoder, and \emph{Olonys}, the nested universal emulator.

\subsection{Encoding databases for analog media}

\begin{figure}[t]
    \centering
    \includegraphics[width=\columnwidth]{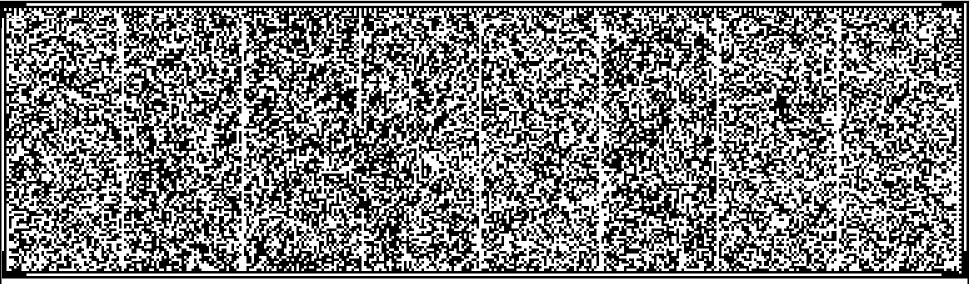}
    \caption{A sample emblem generated by MOCoder from digital data that can be printed to analog media.}
    \label{fig:emblem}
    \vspace{-15pt}
\end{figure}

DBCoder manages compression of archived databases from their textual, software-independent format into a compressed binary layout. Our current DBCoder supports a generic compression scheme based on LZ77 and arithmetic coding that can achieve compression performance close to 7-Zip's LZMA for compressing all database files into a single archive. We are working on supporting more advanced database-specific, compressed, columnar layout schemes as part of on-going research. Irrespective of the layout used, DBCoder is expected to produce a compressed bit stream for further encoding by MOCoder, our media layout coder.

In order to store the compressed bit stream generated by DBCoder using analog media, the bit stream should be converted into visual signals and then printed as pictures. The generated visual signals must be robust to a range of errors that can be introduced during both filming (writing) and scanning (reading). Retrieving digital data stored on film can be jeopardized mainly in two ways. First, the film itself can distort to a small extent over time and become damaged in various ways with fading, hot spots, scratches, etcetera. Second, film scanners are a possible factor of image degradation as they use lenses which can change straight lines into curves, usually near the edge of the field of view. Moreover, as with paper scanners, especially Automatic Document Feeders, often, the mechanical motion in linear array scanners will introduce small perturbations or unsteady movements while scanning.
Dust can also be a source of degradation in microform, both on the film itself, on the glass plates used to hold film while scanning, and also on the surface being filmed which is usually a flat screen with modern microform writers. 


One possible approach to overcoming these problems is to use two-dimensional barcodes, like QR codes and Data Matrix to convert bit streams into barcodes. QR codes, for instance, represent a sequence of bits as an ordered sequence of pixels in a square grid, with white and black pixels representing bits 1 and 0. In addition to these data cells, QR codes also contain a bidimensional clocking system that takes the form of guiding marks placed at specific predefined locations within the field of black and white dots to compensate for distortions. In particular, a QR code always contains 3 position patterns (at 3 corners of the QR code), 2 timing patterns (one for each dimension), and a varying number of alignment patterns, depending on the size of the QR code. These support the decoding algorithm that recovers data back from the QR code by keeping the data bits synchronized with the black and white dots. However, QR codes have been designed considering large-scale distortions, such as an indirect viewing angle using a smartphone camera. Barcodes designed for data archival, in contrast, must be able to cope with low-scale distortions incurred by lenses and unsteady movements of scanners. QR codes are also designed based on the assumption that each dot that makes up the QR code is captured using many pixels, that is, the capture resolution is significantly higher than the QR code resolution. Thus, QR codes and other 2D barcodes typically store a few kilobytes of information at best, and are mainly used as tags or placeholders for short textual information. Archival barcodes, in contrast, should be designed to store multi-megabyte data streams spread over many barcodes.



MOCoder is the media layout encoder used in \systemname{} that performs the ``physical'' layout of bits across barcodes on visual analog media. We refer to the barcodes generated by MOCoder as \emph{emblems} to distinguish them from traditional 2D barcodes. Similar to other 2D codes, MOCoder maps bits to pixels. However, unlike other 2D codes, MOCoder does not use a separate clocking system. Instead, it pairs the bit signal and the clock signal in an approach similar to Differential Manchester encoding used in floppy disks, to generate a self-synchronizing data stream. This approach ensures robust, local clock recovery without having to rely on an independent reference clocking system that could itself be affected by a different level of distortion, as is the case with other 2D barcode schemes. Further, the area that represents data bits is surrounded in each emblem by a thick black square and large-scale black and white dots that allow fast and robust initial detection of the emblem geometry and type, and therefore to precisely position the decoding algorithm on the data area in the scanned image as shown in Figure~\ref{fig:emblem}. 

On top of the physical, visual representation of bits, MOCoder also uses a Reed-Solomon code-based error correcting mechanism to deal with bit erasures that could be caused by media degradation or dust. In particular, MOCoder uses bidimensional error correction with nested Reed-Solomon (RS) codes. The inner RS code works on blocks of data, each holding 223 bytes of user data and 32 redundancy bytes, spread over the entire emblem. This intra-emblem mechanism will automatically correct up to 7.2\% damaged data within a single emblem.
The outer code, or inter-emblem mechanism, protects against whole-emblem failures, by including three parity emblems with each set of 17 data emblems. This results in the full bit-for-bit restoration of data contained within a series of 20 emblems in which any three are missing altogether. 


\begin{table}[t]
    \centering
    \begin{tabular}{|l|l|l|}
    \hline 
    Arithmetic & Logical & Control/Data \\
    \hline 
    ADC(carry) Rd, Rs &  AND Rd, Rs & MOVE Rd, Rs \\
    SBB(borrow) Rd, Rs & OR Rd, Rs  & LDI Rd, \#imm  \\
    SUB Rd, Rs  & XOR Rd, Rs  & LDM Rd, [Ds] \\
    CMP Rd, Rs & LSL/LSR/ASR Rd, Rs  & STM Rs, [Dd] \\
    MUL Rd, Rs & ROR Rd, Rs & JUMP address \\
    \hline 
    \end{tabular}
    \caption{DynaRisc instructions. Rd and Rs refer to destination and source data registers. [Dx] refers to source or destination memory pointer register. \#imm refers to an immediate value.}
    \label{tab:DynaRisc-inst}
    \vspace{-15pt}
\end{table}

\subsection{Olonys: Universal Emulator}

So far, we described the encoding components of DBCoder and MOCoder that transform data from a textual format to emblems. As these encoders are intended to be used by enterprises today, they are written using a contemporary programming language (C\#) and designed to work on a standard Windows computer. Unlike the encoding parts, the decoding parts of DBCoder and MOCoder will be executed several decades in the future to convert data back from emblems into the textual format. Olonys is the universal emulator in \systemname{} that is responsible for archiving these decoders. 

Olonys simulates a simplified, 16-bit, 23-ISA RISC software processor, which we refer to as \emph{DynaRisc}. Table~\ref{tab:DynaRisc-inst} provides a sample of various arithmetic, logical, control transfer and data movement instructions supported by DynaRisc. Further information about the register file, instruction/operand formats, and addressing modes are available in a prior patent publication about Olonys~\cite{olonys-patent}. The key take away from Table~\ref{tab:DynaRisc-inst} is that DynaRisc supports several instructions that are also provided by modern processors. Unlike the encoding part, the decoding part of DBCoder and MOCoder are implemented in DynaRisc assembly. In order to execute the decoders in the future, a user would need to write an emulator that can, in effect, interpret each instruction. Given the limited ISA, writing an emulator for DynaRisc is a much simpler endeavor than writing, for instance, an x86 emulator. However, in order to minimize the amount of work that must be done in the future, and to simplify the task of writing an emulator, Olonys adopts a novel nested emulation strategy. 

Instead of emulating just DynaRisc, Olonys internally emulates two ISAs, DynaRisc and an even-simplified, four-ISA software processor we refer to as VeRisc. The four instructions in the VeRisc ISA are (i) LD \&address (load from memory to general-purpose register $R$), (ii) ST \&address (store from register $R$ to memory), (iii) SBB \&address (subtract from register $R$ the value located at given address), and (iv) AND \&address (logical AND of register $R$ with value at given address). Using just these four VeRisc instructions, we have built an emulator that can interpret the broader DynaRisc ISA. A user now only has to write an emulator for VeRisc, which is effectively implementing an interpreter for just four basic instructions using any computing environment. The VeRisc emulator running on a host computer can then load the DynaRisc emulator written in VeRisc and use it to instantiate the DynaRisc emulator, which, in turn, can load the instruction stream of the decoders written in DynaRisc and instantiate them. Thus, the nested emulation strategy minimizes user effort at restoration time.

Another important design aspect of Olonys is the approach used for bootstrapping the nested emulator. As described earlier, the decoder parts of both DBCoder and MOCoder are implemented using DynaRisc so that Olonys can emulate them, and Olonys internally uses a DynaRisc emulator implemented using VeRisc for executing these decoders. Thus, in order to adhere to the ULE philosophy of storing decoders with the data, we have to store (i) the binary instruction streams of the two decoders, (ii) the binary instruction stream of the DynaRisc emulator, and (iii) description of the VeRisc emulator together with the archived database on analog media. As described earlier, the database itself will be stored in the form of emblems that are generated by MOCoder. Interestingly, we can also convert the DynaRisc instruction stream of DBCoder into emblems and store it similarly to data. This allows for high-density storage of database layout decoding algorithms. Unlike DBCoder, although MOCoder is also programmed using DynaRisc, it cannot be stored as emblems, as it itself is the media layout decoder responsible for decoding and reverse translating emblems into binary. Similarly, the DynaRisc emulator can also not be stored as emblems as the emulator is needed to execute MOCoder before emblems can be converted back into binary data. For this reason, we convert the binary, VeRisc instruction stream corresponding to MOCoder and DynaRisc emulators into a list of textual characters using a text encoding where letters A to P are used to encode hexadecimal values $0xF$ to $0x0$ respectively. This list of characters is stored together with a plain-text description of the VeRisc emulation algorithm that includes the text-decoding logic necessary for converting back the characters into binary values. The result of this procedure is that the entire software stack necessary for decoding data in the future is converted into a short, seven-page document that contains four pages of algorithm pseudocode, and three pages of alphabetic characters, that can be written to analog media with the DBCoder emblems and the database emblems. We refer to this document as the \emph{Bootstrap}. 

\subsection{\systemname{}: Putting it all together}

\begin{figure*}[t]
  \subfloat[][]{
      \includegraphics[width=\columnwidth,page=2]{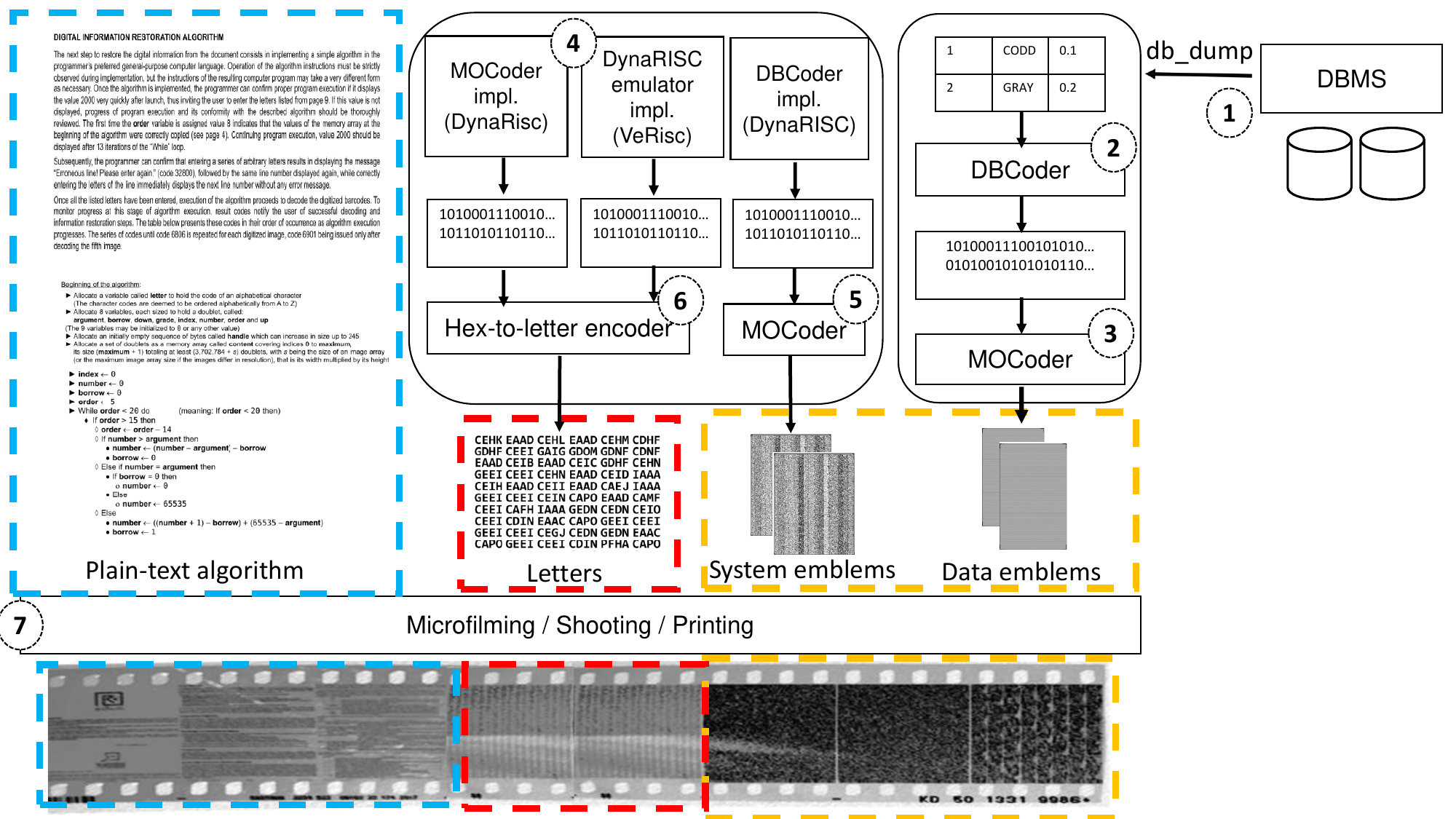}
    }
  \subfloat[][]{
      \includegraphics[width=\columnwidth,page=3]{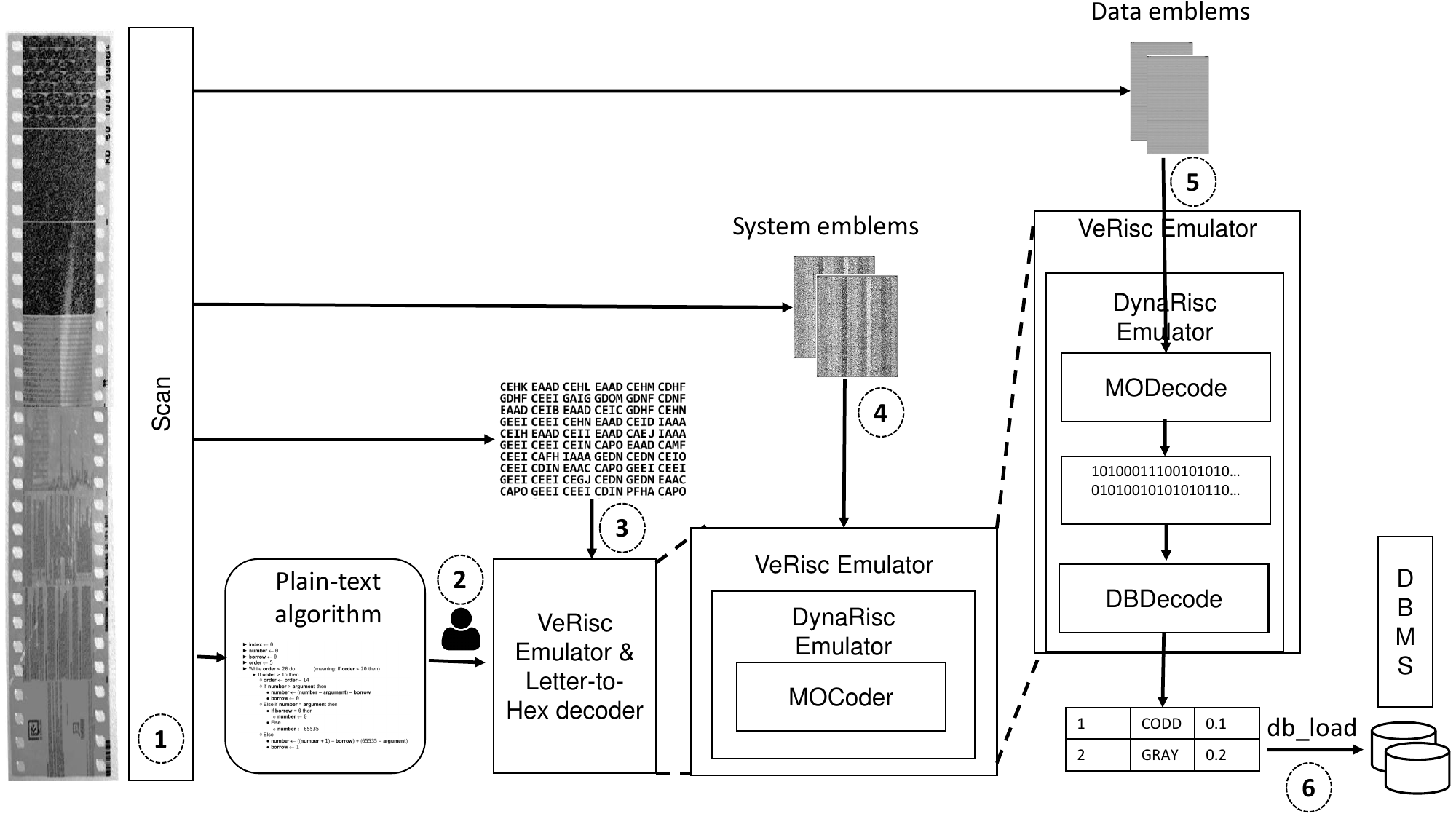}
    }    
  \caption{End-to-end use case; (a) Steps involved in ULE approach of encoding a database together with associated layout decoders using cinema film. (b) Steps involved in decoding and retrieving data back from the film archive.}
  \label{fig:overview}
  \vspace{-10pt}
\end{figure*}

So far, we described the internals of the three major ULE components in \systemname{}. In this section, we will provide an end-to-end usage overview and describe the interaction between various components.

\noindent
\textbf{Archival.} Figure~\ref{fig:overview}(a) shows the internal steps involved in archiving a database using \systemname{}. In the first step, existing database tools are used to extract data out of a database for archival. In the second step, DBCoder is used to convert this data into a compact binary form. The third step shown in Figure~\ref{fig:overview}(a) takes the binary data from DBCoder and uses MOCoder to convert it into \emph{emblems}. The output from MOCoder is a series of high-resolution images that we refer to as \emph{data emblems}. So far, we have described the steps involved in archiving the data. The second column in Figure~\ref{fig:overview}(a) shows the steps involved in archiving the decoders. The fourth step in the overall procedure is writing the decoding parts of DBCoder and MOCoder using DynaRisc. The result of this step are instructions corresponding to these decoders. In the fifth step, the DBCoder DynaRisc instruction stream is passed to MOCoder to generate a new set of  emblems, which we refer to as \emph{system emblems}, to distinguish them from data emblems. Thus, the DBCoder is itself stored as emblems on analog media. 

The sixth step is the archival of MOCoder and the DynaRisc emulator. As described earlier, the DynaRisc instructions of MOCoder and VeRisc instructions of the DynaRisc emulator are converted into a list of letters. These letters are appended to a simple, plain-text pseudocode of the VeRisc emulator to form the Bootstrap. Finally, in the seventh step, the data and system emblems together with the Bootstrap text are physically ``written'' to the analog media via microfilming (for microfilm), shooting (for cinema film), or printing (for archival paper). Figure~\ref{fig:overview}(a) shows a real cinema film that was generated using this approach.   
While we have described all the steps in Figure~\ref{fig:overview}, it is important to note that archival is very simple from the user's point of view. The programming of decoders using DynaRisc and generation of the Bootstrap text is a one-time procedure that is performed in advance by the Olonys developers without any user involvement. 


\noindent
\textbf{Restoration.} Figure~\ref{fig:overview}(b) shows the steps involved in restoring the database using \systemname{}. The first step in restoring the data is scanning the microform to generate high-resolution images corresponding to each frame. The user extracts the three pages of alphabetical characters
that correspond to the instruction streams of the DynaRisc emulator and MOCoder from the images. Any OCR program can be used to automate this task. Similarly, the user converts the images containing emblems into a linear flat array of pixel intensities as described in the Bootstrap. Any standard image handling libraries can be used for automating this task. With image preprocessing done, in the second step, the user implements the VeRisc emulator using the algorithm pseudocode in the Bootstrap. It is important to note here that we make no assumptions about the underlying hardware or software environment on which the emulator will run. The pseudocode is less than 500 lines of code that can be implemented by anyone with a basic programming background. The user then executes the code, thereby emulating a basic, four-instruction virtual machine. In the third step, the Bootstrap code reads the alphabetical characters, decodes them, and instantiates a new DynaRisc emulator as an application inside the VeRisc environment, and executes MOCoder within the DynaRisc emulator. In the fourth step, the system emblems are then read, decoded by MOCoder, and used to load DBCoder. In the fifth step, the remaining data emblems are then read, and decoded by MOCoder first and then by DBCoder to output the ASCII database archive files. Finally, in the sixth and final step, the user loads the standardized data into any future DBMS using then-current tools and interfaces.

Notice that unlike archival, all the steps during restoration are performed by the user without our involvement. \systemname{} is deliberately designed in such a way that the most complex step, the implementation of the VeRisc emulator, is as simple as possible so that the user will still be able to recover the data even if the creators of the system are no longer alive. The Bootstrap provides technical instructions to precisely guide the user through the process of setting up the environment necessary for decoding data.

\vspace{-10pt}



\section{Evaluation}

To demonstrate the feasibility of ULE, we performed several end-to-end experiments where we used \systemname{} to archive and restore data using analog media. In this section, we provide details about our experiments that show that the ULE approach can be realized in practice, and analog media can indeed be integrated into the database storage hierarchy.

\noindent
\textbf{Paper archive.} For our first experiment, we used the industry-standard TPC-H benchmark to generate a test dataset. We loaded the data into a PostgreSQL database and used pg\_dump to generate the database archive in the text-based SQL format. We configured the TPC-H scale factor to produce an archive file that was roughly 1MB in size (1.2MB). We used \systemname{} to encode this archive into 26 emblems that were directly printed to A4 paper at 600 dpi using a network-attached Canon ImageRunner Advance 6255i Laser printer. Thus, we achieved a density of 50KB per page. Replacing our A4 paper with an archival-grade one would be the only change required for archiving a database to permanent paper. The combined encoding and printing process took 6 minutes on a low-end Windows laptop equipped with Intel Core i7-6500U CPU clocked at 2.5GHz, and 16GB of DRAM. 
In order to test our decoding process with a different computing platform locally, 
we used the same equipment to scan the emblems back as 26 pdf files. We then implemented the VeRisc emulator in C++ and executed it on a Linux server equipped with an Intel Core i9-10920X CPU clocked at 3.5 GHz. The decoding process successfully restored back the SQL archive file in 3 minutes and 20 seconds.

\noindent
\textbf{Microfilm archive.} In order to show that \systemname{} also works with microform, we targeted a 16mm microfilm as the archival media for our second experiment. We used the EPM/Kodak IMAGELINK 9600 Archive Writer\footnote{https://www.epminc.com/products/microfilm-equipment/imagelink-9600/}  for ``writing'' to microfilm. With this equipment, each frame written to film is a 3888 (width) x 5498  (length) pixel black and white (bitonal) TIFF image. With such a system, \systemname{} is capable of storing 1.3GB in a single 66 meter reel. Due to time and budget constraints, we were able to use \systemname{} to only encode a 102KB TIFF image (the Olonys logo), instead of the 1MB PostgreSQL database. The image was encoded into 3 emblems by \systemname{}, and these emblems were written together with the Bootstrap to the 16mm microfilm. A standard microfilm reader\footnote{https://history.denverlibrary.org/news/wait-minute-you-still-use-microfilm}  was used to scan back the emblems. The produced scans were also bitonal with a high resolution of about 5000 x 7000 pixels. We used our VeRisc emulator to convert the emblems back to the source image without any errors. 

\noindent
\textbf{Cinema film archive.} A similar experiment was conducted with 35mm black and white cinema film shown in Figure~\ref{fig:overview}. The same Olonys logo was ``shot'' as 3 emblems in 3 full-aperture frames (equivalent to the 4/3 image ratio) with a resolution of 2048 x 1556 pixels (2K) using the Arrilaser digital film recorder. The frames were then scanned in grayscale 4K resolution (4096 x 3120 pixels) using a Scanity Immersion from DFT. Both shooting and scanning use the specific DPX image format used for raw cinema frames. Compared to microfilm scanners, we found cinema film scanners to produce sharper, low-distortion images. We used our VeRisc emulator to convert these emblems back to the source image successfully. 

While we are pursuing large-scale database archival experiments as a part of on-going research, we believe that our preliminary results with the 102KB image demonstrate clearly that \systemname{} can work with any visual analog backend. 

\noindent
\textbf{Portability and user friendliness.} The aforementioned experiments demonstrated that the ULE approach can be realized in practice, and \systemname{} can successfully archive data to analog media. However, in order to ensure that a user in the distant future can implement the VeRisc emulator on a computing platform that is unknown today, we also undertook two additional tasks. First, we requested people with diverse technical backgrounds, including first-year undergraduate students (at Lyc\'{e}e Bonaparte, Toulon), engineers at a partner institute (CNES), and researcher staff at EURECOM, to implement the VeRisc emulator in any language and system of their choice. Thus, the emulator was implemented on Windows and Linux in several programming languages including JavaScript, Python, C++, and C\#. In all cases, we found that our ``users'' were able to implement a fully functional emulator without any prior knowledge of the system in under a week. Second, we have also ported the Olonys universal emulator to several computing platforms that use various microprocessors, including Raspberry Pi and GameBoy Advance (ARM), TI-85 calculator (Z80), Atari Falcon (68030), and Palm PDA (68000). In doing so, we ensured that there were no architecture-specific aspects in Olonys that would hinder its implementation on future platforms. 
\section{Conclusion}

With the growing adoption of data-driven decision making, enterprises are increasingly facing the need to archive data over long time periods to meet legal and regulatory compliance requirements. 
In this paper, we introduced Universal Layout Emulation as a new approach for long-term database archival that uses universal emulation for archiving layout decoders together with the data. In order to show that ULE can be realized in practice, we presented \systemname{}, an end-to-end long-term data archival system based on ULE. Using an experimental evaluation, we demonstrated that \systemname{} is portable, easy to use, and can be used to archive databases using analog media. 

In future work, we plan to extend \systemname{} on several fronts. First, we are working on adding support for compressed, columnar layout encoding schemes in DBCoder that are well-known to provide an order of magnitude reduction to storage utilization over the generic compression support available today. Second, despite its longevity, analog media might not be suited for extremely large databases due to density issues. For example, \systemname{} can store 1.3GB in a 66m microfilm reel. This implies that one would need 800 reels for Terabyte-sized data lakes, and hundreds of thousands of reels for Petabyte-sized data lakes. Thus, while microfilm might be a feasible solution for small or medium-sized archives, it is unsuitable for extremely large archives. DNA, in contrast has a theoretical density of 1EB per $mm^3$. Thus, one avenue of future work we are pursuing is extending \systemname{} to be used in conjunction with a DNA-based database archive~\cite{oligoarchive}. 



\bibliographystyle{ACM-Reference-Format}
\bibliography{sample}

\end{document}